\documentclass[prb,twocolumn,letterpaper]{revtex4}
\usepackage{dcolumn}
\usepackage{graphicx}

\begin{document}

\bibliographystyle{apsrev}

\title{Some Considerations on the Additional Absorption Peak  
in the $c$-axis Infrared Conductivity of Bilayer Cuprate Superconductors:
Interpretation of the Changes Induced by a Parallel Magnetic Field 
and the Role of Bilayer Splitting} 

\author{Dominik Munzar} 

\affiliation{Institute of Condensed Matter
Physics, Faculty of Science, Masaryk University,
Kotl\'{a}\v{r}sk\'{a} 2, 61137 Brno, Czech Republic}

\begin{abstract}
Changes of the $400{\rm\,cm^{-1}}$ peak 
in the $c$-axis conductivity of underdoped YBa$_{2}$Cu$_{3}$O$_{6.6}$ 
upon application of a parallel magnetic field 
reported by Kojima {\it et al.}
are shown to be consistent 
with the model where the peak is due to the superfluid. 
Results of our calculations 
of the $c$-axis response of bilayer compounds 
with well defined bilayer-split bands 
are presented and discussed. 
For moderate values of the bilayer splitting 
($\Delta\epsilon$ comparable to $2\Delta_{\rm max}$)
the spectra of the superconducting state   
exhibit an additional mode which is due to the condensate  
and similar to the one of earlier phenonomenological approaches. 
\end{abstract}

\maketitle

The $c$-axis infrared spectra 
of bilayer high-$T_{c}$ cuprate superconductors (HTCS),
i.e., HTC compounds with two CuO$_{2}$ planes 
within a unit cell, 
exhibit the following interesting features. 
At low temperatures a broad absorption peak develops 
in the spectra of the real part of the optical conductivity $\sigma_{1}$ 
in the frequency region  
between  $300{\rm\,cm^{-1}}$ and $800{\rm\,cm^{-1}}$ 
(its frequency is material- and doping-dependent) 
\cite{Homes,Schutzman,Bernhard,Zelezny,Dordevic}. 
At the same time some of the phonon modes 
are strongly renormalized, i.e., their frequencies and/or linewidths
and/or spectral weights change
\cite{Homes,Schutzman,Bernhard,Zelezny}. 
The magnitude of the additional peak and the changes of the phonons 
are particularly spectacular for the strongly underdoped 
superconductor YBa$_{2}$Cu$_{3}$O$_{7-\delta}$ (Y-123) 
\cite{Munzar,Bernhard}.  
For a long time these phenomena had 
retained their mystery.   
In 2000 Gr{\"u}ninger {\it et al.}~\cite{Gruninger} 
showed that the additional peak in the spectra  
of Y-123 
can be understood and quantitatively described 
in terms of a model proposed in 1996 
by Van der Marel and Tsvetkov \cite{VdM}. 
Within this model, the so called multilayer model,  
a bilayer compound is considered as a superlattice 
of homogeneously charged superconducting planes 
and interbilayer and intrabilayer regions as shown in Fig.~1.  

\begin{figure}[htbp]
  \centering
  \includegraphics[width=0.4\textwidth]{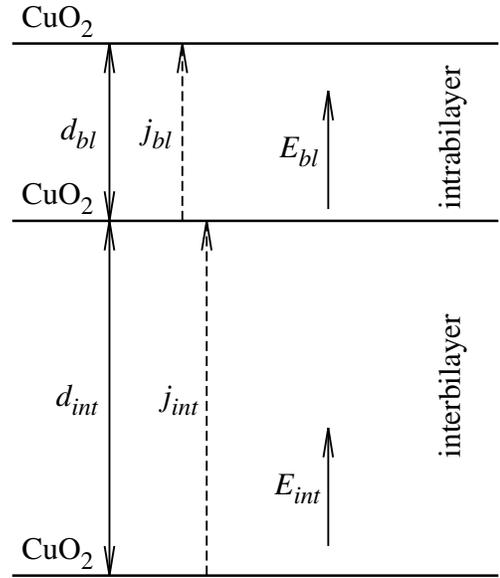}  
  \caption{Schematic representation of the multilayer model.} 
\end{figure}

The model dielectric function $\varepsilon(\omega)$ is given by 
$$
{d_{bl}+d_{int}\over \varepsilon(\omega)}=
{d_{bl}\over \varepsilon_{bl}(\omega)}+
{d_{int}\over \varepsilon_{int}(\omega)}\,,
\eqno (1)
$$
$$
\varepsilon_{bl/int}(\omega)=\varepsilon_{\infty}
+{i\sigma_{bl/int}(\omega)\over \varepsilon_{0}\omega}\,,
\eqno (2) 
$$
where $\varepsilon_{\infty}$ is the interband dielectric constant and 
the conductivities $\sigma_{bl}$ and $\sigma_{int}$  
are defined by $j_{bl}=\sigma_{bl}E_{bl}$ and $j_{int}=\sigma_{int}E_{int}$.
In the superconducting state (SCS), 
they contain contributions of the condensate, 
$$
\sigma_{bl/int}(\omega)={i\varepsilon_{0}\omega_{bl/int}^{2}\over \omega}
+\sigma_{bl/int}^{reg}(\omega)\,,
\eqno (3) 
$$
where $\omega_{bl}$ and $\omega_{int}$ 
are the plasma frequencies.
Further $\sigma_{bl}^{reg}$ and $\sigma_{int}^{reg}$ are 
the regular parts of the conductivities. 
In the absence of the latter, 
$\varepsilon(\omega)$ 
exhibits two zero crossings, at   
$\omega_{L1}=\sqrt{\omega_{int}/\varepsilon_{\infty}}$ and at 
$\omega_{L2}=\sqrt{\omega_{bl}/\varepsilon_{\infty}}$, 
corresponding to two longitudinal plasma modes: 
the interbilayer and the intrabilayer one;  
in addition, it has a pole at 
$$
\omega_{T}=\sqrt{{d_{bl}\omega_{int}^{2}+d_{int}\omega_{bl}^{2}\over
(d_{bl}+d_{int})\varepsilon_{\infty}}}
\eqno (4) 
$$
corresponding to a so called transverse plasmon (TP).  
Its spectral weight is given by 
$$
S_{T}={\pi\over 2}\varepsilon_{0}
{d_{bl}d_{int}(\omega_{bl}^{2}-\omega_{int}^{2})^{2}\over 
(d_{bl}+d_{int})(d_{bl}\omega_{int}^{2}+d_{int}\omega_{bl}^{2})}\,.
\eqno (5)
$$
Gr{\"u}ninger {\it et al.}  
attributed the additional peak to the TP. 
The model has then been extended \cite{Munzar}
by including local fields acting on the ions.
This allowed the authors of Ref.~\onlinecite{Munzar} 
to explain not only the additional peak 
but also the related phonon anomalies. 
Today, the multilayer model \cite{VdM} 
and the local field idea \cite{Munzar}
are becoming accepted. 
The nature of the intra-bilayer currents 
responsible for the additional peak, however, 
is the subject of an ongoing discussion  
\cite{Timusk,Dordevic}.
Here we address two related issues. 

{\it Changes of the additional peak in strongly underdoped Y-123 
induced by a parallel magnetic field} 
have been studied by Kojima {\it et al.} \cite{Kojima1}. 
It has been found that the field suppresses the condensate at $\omega=0$. 
Its spectral weight (SW) 
is redistributed between a so called $\sigma$ mode at ca $40{\rm\,cm^{-1}}$ 
and the additional peak at ca $400{\rm\,cm^{-1}}$. 
The SW increase of the latter amounts to ca ${460\,\Omega^{-1}{\rm cm}^{-2}}$. 
Surprisingly, the field does not influence the peak frequency. 
Below we show that this behaviour is consistent 
with the claim that the peak corresponds to the superconducting condensate. 
Assume for simplicity 
that the suppression of the superfluid in the interbilayer regions is complete, 
i.e., $\omega_{int}=0$, 
that the field does not modify the intrabilayer plasma frequency $\omega_{bl}$ 
( $\omega_{bl}>>\omega_{int}$), 
and that the regular parts of the two conductivities can be neglected.
The field induced changes of $S_{T}$ and $\omega_{T}$ 
can then be obtained using Eq.~(5) and Eq.~(4), respectively,  
$$
{\Delta S_{T}\over S_{T}}\approx 
{(2d_{int}+d_{bl})\over d_{int}}
{\omega_{int}^{2}\over \omega_{bl}^{2}}\,,
\eqno (6)
$$
$$
{\Delta \omega_{T}\over \omega_{T}}\approx 
-{d_{bl}\over 2d_{int}}
{\omega_{int}^{2}\over \omega_{bl}^{2}}\,.
\eqno (7) 
$$
It appears that 
$S_{T}$ somewhat increases 
and $\omega_{T}$ very slightly decreases. 
For the values of the parameters
obtained by fitting the spectra of a similarly underdoped Y-123 sample, 
$\omega_{bl}=1200{\rm\,cm^{-1}}$, $\omega_{int}=220{\rm\,cm^{-1}}$ 
(Ref.~\onlinecite{Munzar}),  
$d_{bl}=3.3$ {\AA} and $d_{int}=8.4$ {\AA},
we obtain 
$\Delta S_{T}=800{\,\Omega^{-1}{\rm cm}^{-2}}$ 
and $\Delta \omega_{T}= -3 {\rm\,cm^{-1}}$. 
In reality, the suppression of the condensate in the interbilayer regions 
is certainly incomplete. 
Kojima and coworkers successfully interpreted 
the low-frequency part of the spectra 
in terms of a model where the Josephson coupling is considerably weakened 
only in ca $50\%$ of the interbilayer regions.  
Using this model we obtain 
$\Delta S_{T}\approx 400{\,\Omega^{-1}{\rm cm^{-2}}}$ 
and $\Delta \omega_{T}\approx -1.5{\rm\,cm^{-1}}$ 
[the magnitudes are reduced ca by a factor of 0.5 
with respect to the simple estimates of Eqs.~(6) and (7)] \cite{Marek}.  
The value of $\Delta S_{T}$ is in excellent agreement with experiment.  
The very small value of $\Delta \omega_{T}$ 
accounts for the fact that, at first glance, 
the $400{\rm\,cm^{-1}}$ peak does not shift upon application of the field. 

{\it Transverse plasmon in the presence of bilayer splitting.} 
The multilayer model of Eq.~(1)
is well justified \cite{Shah}
for a system of weakly (``Josephson") coupled CuO$_{2}$ planes  
where $t_{\perp\,{\rm max}}$ is much less 
than any important in-plane energy scale, 
in particular $t_{\perp\,{\rm max}}<\Delta_{\rm max}$.  
Here $t_{\perp\,{\rm max}}$ and $\Delta_{\rm max}$ 
are the maximum values of the interplanar hopping matrix element 
and the superconducting gap, respectively. 
The overdoped bilayer compound Bi$_{2}$Sr$_{2}$CaCu$_{2}$O$_{8}$ (Bi-2212), 
however,  
exhibits two bands: the bonding band (BB) and the antibonding band (AB),   
with the maximum splitting of ca $90{\rm\,meV}$ 
corresponding to $t_{\perp\,{\rm max}}=45{\rm\,meV}>\Delta_{\rm max}$ 
\cite{Feng,Chuang}.
Here we present very preliminary results of our calculations 
of the $c$-axis response in such a system.  
Details will be presented in Ref.~\onlinecite{Munzarc}. 
The main ingredients of our approach are summarized below. 

Our starting point is the tight-binding formalism  
outlined in Appendix B of Ref.~\onlinecite{ILT}
(with a more general form of the interplanar hopping matrix elements). 
For simplicity, the spacing layers will be supposed to be fully insulating 
(i.e., $t_{\perp\,int}=0$ in the notation of Ref.~\onlinecite{ILT}). 
First, we express the intrabilayer conductivity $\sigma_{bl}(\omega)$ defined 
by $j_{bl}(\omega)=\sigma_{bl}(\omega)E_{bl}(\omega)$. 
It is given by Eq.~(B9) of Ref.~\onlinecite{ILT}:
$$
\sigma_{bl}(\omega)={[e^{2}d_{bl}/(a^{2}\hbar^{2})]
\langle T\rangle+\chi({\mathbf q}=0,\omega)\over i(\omega+i\delta)}\,,
\eqno (8)
$$
where $a$ is the in-plane lattice constant, 
$T$ is the intrabilayer $c$-axis kinetic energy per unit cell, 
$$
\langle T\rangle=-{2a^{2}\over (2\pi)^{2}}
\int_{2D\,BZ}{\rm d}{\mathbf k}\, t_{\perp}({\mathbf k})
(n_{Bk}-n_{Ak})\,,
\eqno (9)
$$
and $\chi$ 
is the relevant current-current correlation function, 
$$
\chi({\mathbf q},\omega)={Na^{2}d_{bl}i\over \hbar}
\int_{-\infty}^{\infty}{\rm d}t
\langle [j({\mathbf q},t),j(-{\mathbf q},0)]
\rangle 
\Theta(t)e^{i\omega t}\,.
\eqno (10)
$$
The interplanar hopping matrix element $t_{\perp}$  
is related to the dispersion relations of the two bands as
$
2t_{\perp}({\mathbf k})=
\epsilon_{Ak}-\epsilon_{Bk}\,.
$
The occupation factors are denoted by 
$n_{Bk}$ and $n_{Ak}$. 
In Eq.~(10), $j({\mathbf q},t)$ is the Fourier component 
of the paramagnetic part of the intrabilayer current density operator.
It can be shown \cite{Munzarc} that 
$$
j({\mathbf q}=0)={ie\over Na^{2}\hbar}
\sum_{{\mathbf K}=({\mathbf k},k_{z})}t_{\perp}({\mathbf k})
(c_{BK}^{+}c_{AK}-c_{AK}^{+}c_{BK})\,,
\eqno (11)
$$
where $c_{BK}^{+}$, $c_{BK}$, $c_{AK}^{+}$, and $c_{AK}$ 
are the conventional quasiparticle operators. 
The Matsubara counterpart $\chi(i\omega_{m})$ 
of $\chi({\mathbf q}=0,\omega)$ is given by \cite{Munzarc}
$$
\chi(i\omega_{m})={4e^{2}d_{bl}\over \hbar^{3}}
{1\over (2\pi)^{2}}
\int\,{\rm d}{\mathbf k}\, t_{\perp}^{2}({\mathbf k}) \times
$$
$$
\times 
\left\{
-{1\over 2\beta}\sum_{ip_{n}}{\rm Tr}
\left[
G_{B}({\mathbf k},ip_{n})G_{A}({\mathbf k},ip_{n}+i\omega_{m})
\right]
\right\}\,,   
\eqno (12)
$$
where $G_{B}$ and $G_{A}$ are the Nambu propagators. 
Note that for $G_{B}=G_{A}$ 
(i.e., in the limit of weak interlayer coupling) 
we recover Eq.~(10) of Ref.~\onlinecite{Shah}.  
When deriving Eq.~(12), vertex corrections have been neglected. 
In order to express the dielectric function,  
we have to consider interlayer Coulomb interactions.
Random phase approximation leads to Eqs.~(1), (2) with $\sigma_{int}=0$. 
The multilayer formula (1) is thus justified 
even if the coupling between the closely-spaced planes is strong 
(the coupling across the spacing layer, however, must be negligible). 

The results presented below have been obtained using the above equations
with selfenergy corrections neglected. 
The final formula for $\chi$ reads 
$$
\chi(\omega)={e^{2}d_{bl}\over \hbar^{2}(2\pi)^{2}}
\int\,{\rm d}{\mathbf k}\, 
t_{\perp}^{2}({\mathbf k})\times
$$
$$
\times 
\biggl(
l_{1}({\mathbf k})[1-n_{F}(E_{Bk})-n_{F}(E_{Ak})]\times 
$$
$$
\times 
\left\{
{1\over \hbar\omega+i\delta-E^{+}_{k}}-
{1\over \hbar\omega+i\delta+E^{+}_{k}}
\right\}
$$
$$
+l_{2}({\mathbf k})[n_{F}(E_{Ak})-n_{F}(E_{Bk})]\times
$$
$$
\times
\left\{
{1\over \hbar\omega+i\delta-E^{-}_{k}}-
{1\over \hbar\omega+i\delta+E^{-}_{k}}
\right\}
\biggr)\,,
\eqno (13)
$$
where 
$$
l_{1/2}({\mathbf k})=
{\epsilon_{Bk}\epsilon_{Ak}+\Delta_{Bk}\Delta_{Ak}\mp E_{Bk}E_{Ak}\over 
E_{Bk}E_{Ak}}\,,
\eqno (14)
$$
$\Delta_{Bk/Ak}$ are the superconducting gaps, 
$E_{Bk/Ak}=\sqrt{\epsilon_{Bk/Ak}^{2}+\Delta_{Bk/Ak}^{2}}$, 
$n_{F}$ is the Fermi function and  
$E^{\pm}_{k}=E_{Ak}\pm E_{Bk}$. 
A conventional form of the dispersion relations \cite{OKA} has been used: 
$$
\epsilon_{B/A}(k)=-2t[\cos(k_{x}a)+\cos(k_{y}a)]-
$$
$$
4t'\cos(k_{x}a)\cos(k_{y}a)
\mp 
{t_{\perp\,max}\over 4}[\cos(k_{x}a)-\cos(k_{y}a)]^{2}-\mu\,,
\eqno (15)
$$
and the bands have been assumed to possess the same superconducting gap
$$
\Delta_{Bk}=\Delta_{Ak}={\Delta_{max}\over 2}[\cos(k_{x}a)-\cos(k_{y}a)]\,.
\eqno (16)
$$
 
Figure 2 shows the main results 

\begin{figure}[htbp]
  \centering
  \includegraphics[width=0.5\textwidth]{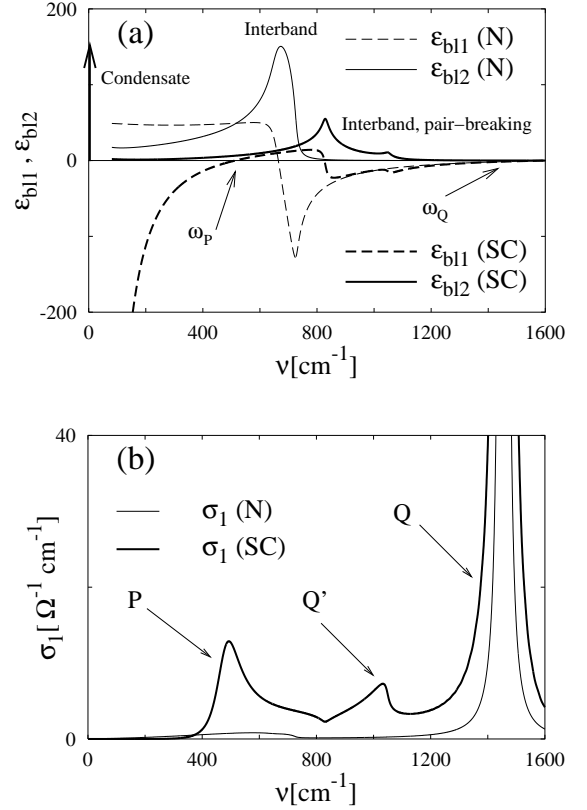}  
  \caption{Calculated spectra 
of the real ($\varepsilon_{bl1}$) 
and imaginary ($\varepsilon_{bl2}$) parts 
of the ``intrabilayer dielectric function" $\varepsilon_{bl}$ 
defined by Eq.~(2) (a) 
and of the real part of the $c$-axis conductivity (b).}
\end{figure}

for the following values of the model parameters:  
$d_{bl}=3.4$ {\AA}, $d_{int}=12.0$ {\AA} 
(the values correspond to Bi-2212), 
$\varepsilon_{\infty}=5.0$, 
$t=250{\rm\,meV}$, 
$t'=-100{\rm\,meV}$, 
$t_{\perp\,{\rm max}}=45{\rm\,meV}$ \cite{Feng}, 
$\mu=-350{\rm\,meV}$,   
$\Delta_{\rm max}=30{\rm\,meV}$, and
$\delta=1.0{\rm\,meV}$.  
In the normal state (NS, $\Delta_{\rm max}=0$, $T=100{\rm\,K}$),  
$\varepsilon_{bl\,2}(\omega)$ contains only the interband (IB) contribution. 
Its maximum is located at 
$\omega_{IB}\approx 2t_{\perp\,{\rm max}}/\hbar$.
In the SCS ($T=20{\rm\,K}$), 
$\varepsilon_{bl\,2}(\omega)$ consists of the IB (pair-breaking) contribution 
centered at 
$\omega_{IB}\approx 2 \sqrt{t_{\perp\,{\rm max}}^{2}+
\Delta_{\rm max}^{2}}/\hbar$, 
with a smaller SW than in the NS, 
and the one of the condensate (C) at zero frequency. 
The ratio ${\rm SW(C)/SW(IB)}$ depends 
on $\Delta_{\rm max}/t_{\perp\,{\rm max}}$.
For $\Delta_{\rm max}=20{\rm meV}$ ($30{\rm meV}$;  $60{\rm meV}$), e.g., 
we obtain 
$\omega_{bl}=1870{\rm\,cm^{-1}}$, ${\rm SW(C)/SW(IB)}=0.47$ 
($2260{\rm\,cm^{-1}}$, $0.93$; 
$2690{\rm\,cm^{-1}}$, $3.07$).  
The conductivity $\sigma_{1}(\omega)$ 
has sharp maxima labelled as P and Q
(we do not consider here the small maximum labelled as Q'). 
They occur at frequencies $\omega_{P}$ and $\omega_{Q}$, for which 
$\varepsilon_{bl\,1}=-(d_{bl}/d_{int})\varepsilon_{\infty}$ 
and $\varepsilon_{bl\,2}$ is small 
[see Eq.~(1)]. 
The maximum Q has a large SW 
of about $60 000{\,\Omega^{-1}{\rm cm}^{-2}}$.
It appears already in the NS
and it will be broadened by selfenergy effects. 
The large difference between $\omega_{Q}$ 
and $\omega_{IB}$  
is caused by the interlayer Coulomb interactions.  
The maximum (mode) P appears only in the SCS 
as a result of broken gauge symmetry. 
Its frequency and SW ($S_{P}$) increase dramatically with 
increasing $\Delta_{\rm max}/t_{\perp\,{\rm max}}$.
For $\Delta_{\rm max}=20{\rm meV}$ ($30{\rm meV}$; $60{\rm meV}$), e.g., 
we obtain 
$\omega_{P}=350{\rm\,cm^{-1}}$, $S_{P}\approx 1300{\,\Omega^{-1}{\rm cm^{-2}}}$
($490{\rm\,cm^{-1}}$, $2500{\,\Omega^{-1}{\rm cm^{-2}}}$;
$840{\rm\,cm^{-1}}$, $13600{\,\Omega^{-1}{\rm cm^{-2}}}$). 
For $\Delta_{\rm max}/t_{\perp\,{\rm max}} \rightarrow 0$ 
(i.e., in the limit of strong interlayer coupling), the mode vanishes. 
Obviously, the maximum P can be related to the TP 
of the phenomenological model 
involving a superlattice of interbilayer and intrabilayer Josephson junctions. 
The underlying excitation can be visualized 
as a resonant oscillation of the condensate density 
between the closely-spaced CuO$_{2}$ planes. 

In conclusion, the changes of the $400{\rm\,cm^{-1}}$ peak 
induced by a parallel mg.~field 
reported by Kojima {\it et al.} 
are consistent with the claim 
that it is related to the superconducting condensate.  
The regular parts of the two conductivities 
need not be taken into account 
in order to explain the changes. 
The $c$-axis response of bilayer compounds
has been studied considering the B-A splitting. 
The conductivity $\sigma_{1}$ exhibits a maximum in MIR 
related to the B-A transitions. 
Its shape will be determined by selfenergy terms 
and it can be expected to change slightly 
when going from the NS to the SCS.  
{\it Only in the SCS}, an additional maximum appears in FIR,  
related to the condensate, 
with similar properties as the one  
of the naive Josephson-superlattice model.
Our results indicate that the latter model is compatible 
with moderate values of the B-A splitting.   

This work has been supported by the Ministry of Education of CR (CEZ 143100002).


\begin{thebibliography}{La86}
\bibitem{Homes} C.~C.~Homes {\it et al.}, 
Physica C {\bf 254}, 265 (1995). 
\bibitem{Schutzman} 
J.~Sch{\"u}tzman {\it et al.}, 
Phys.~Rev.~B {\bf 52}, 13665 (1995).  
\bibitem{Bernhard}
C.~Bernhard {\it et al.}, 
Phys.~Rev.~B 
{\bf 61}, 618 (2000).  
\bibitem{Zelezny}
V.~{\v Z}elezn{\'y} {\it et al.}, 
Phys.~Rev.~B {\bf 63}, 060502 (2001). 
\bibitem{Dordevic} S.~V.~Dordevic {\it et al.}, 
Phys.~Rev.~B {\bf 69}, 094511 (2004). 
\bibitem{Munzar} D.~Munzar {\it et al.}, 
Solid State Commun.~{\bf 112}, 365 (1999).
\bibitem{Gruninger} M.~Gr{\"u}ninger {\it et al.}, 
Phys.~Rev.~Lett.~{\bf 84}, 1575 (2000). 
\bibitem{VdM} D.~van der Marel and A.~Tsvetkov,
Czech.~J.~Phys.~{\bf 46}, 3165 (1996); 
Phys.~Rev.~B {\bf 64}, 024530 (2001). 
\bibitem{Timusk} T.~Timusk and C.~C.~Homes, 
Solid State Commun.~{\bf 126}, 63 (2003). 
\bibitem{Kojima1} 
K.~M.~Kojima {\it et al.}, Phys.~Rev.~Lett.~{\bf 89}, 247001 (2002). 
\bibitem{Marek} J.~Marek and D.~Munzar, to be published.
\bibitem{Shah} N.~Shah and A.~J.~Millis,
Physical Review B {\bf 65}, 024506 (2001).
\bibitem{Feng} D.~L.~Feng {it et al.}, Phys.~Rev.~Lett.~{\bf 86}, 
5550 (2001). 
\bibitem{Chuang} Y.~D.~Chuang {\it et al.}, Phys.~Rev.~Lett.~{\bf 87}, 
117002 (2001). 
\bibitem{Munzarc} D.~Munzar {\it et al.}, to be published. 
\bibitem{ILT} D.~Munzar {\it et al.}, Phys.~Rev.~B {\bf 64}, 024523 (2001). 
\bibitem{OKA} O.~K.~Andersen {\it et al.}, 
J.~Phys.~Chem.~Solids {\bf 56}, 1573 (1995). 
\end{thebibliography}
\end{document}